\documentclass[aps,prl,twocolumn,groupedaddress,showpacs,secnumarabic,amssymb,amsmath,tightenlines]{revtex4}

\usepackage{graphicx}

\begin{document}
\title{Orientation and Motion of Water Molecules at Air/Water Interface}
\author{Wei Gan}
\thanks{Also Graduate School of the Chinese Academy of Sciences}
\author{Dan Wu}
\thanks{Also Graduate School of the Chinese Academy of Sciences}
\author{Zhen Zhang}
\thanks{Also Graduate School of the Chinese Academy of Sciences}
\author{Hong-fei Wang}
\email{hongfei@mrdlab.icas.ac.cn}
\thanks{Author to whom correspondence
should be addressed.} \affiliation{State Key Laboratory of
Molecular Reaction Dynamics,
\\Institute of Chemistry, the Chinese Academy of Sciences,
Beijing, China, 100080}

\date{\today}

\begin{abstract}
Analysis of SFG vibrational spectra of OH stretching bands in four
experimental configurations shows that orientational motion of
water molecule at air/water interface is libratory within a
limited angular range. This picture is significantly different
from the previous conclusion that the interfacial water molecule
orientation varies over a broad range within the vibrational
relaxation time, the only direct experimental evidence for
ultrafast and broad orientational motion of a liquid interface by
Wei \textit{et al.} [Phys. Rev. Lett. 86, 4799, (2001)] using
single SFG experimental configuration.
\end{abstract}


\maketitle

Due to lack of direct experimental measurement techniques, motion
and dynamics of molecules at liquid interface are yet to be well
understood \cite{ShenMirandaJPCBReview}. In the past decade,
air/water interface, as one of the most important liquid
interface, has been intensively investigated both theoretically
and experimentally
\cite{ShenMirandaJPCBReview,Mundy-science,BenjaminPRL1994CR,HynesCP2000JPCB,MooreJCP2003,
FeckoScience2003,ChandraCPL20032004,RiceJCP1991,EisenthalJPC1988,
PershanPRL19851988,SaykallyJPCM2002,DuQuanPRL1993,Richmond-jpca2000,WeiXingPRL2001}.
Experimental method such as X-Ray reflection can be used to get
the liquid surface roughness \cite{PershanPRL19851988}, but
orientational structure and dynamics information at molecule level
can only be obtained by Sum Frequency Generation (SFG) or Second
Harmonic Generation (SHG) because of their submonolayer
sensitivity and interface specificity
\cite{ShenMirandaJPCBReview}. It has been generally accepted that
the interfacial water molecule has one free OH bond protruding out
of the interface. However, the existence of water molecules with
both OH bonds pointing to the vapor phase is still a controversial
issue
\cite{DuQuanPRL1993,Richmond-jpca2000,WeiXingPRL2001,SaykallyJPCM2002,Mundy-science}.
The dynamics of water molecule at air/water interface has been
discussed theoretically in literatures \cite{Mundy-science,
BenjaminPRL1994CR,HynesCP2000JPCB,ChandraCPL20032004,MooreJCP2003,
RiceJCP1991}. The only direct experimental study so far concluded
a fast orientational motion varies over a broad range of
$102^{\circ}$ centered at the surface normal
($\theta_{M}=51^{\circ}$) within a time scale comparable or less
than 0.5 \textit{ps}, by Wei \textit{et al.} from the fact of the
vanishing of the SFG vibrational spectra of the free OH stretching
mode of the interfacial water molecule around $\sim 3700cm^{-1}$
in the \textit{sps} polarization (denoting \textit{s}-
\textit{p}-,  and \textit{s}- polarized sum frequency output,
visible input, and infrared input, respectively)
\cite{WeiXingPRL2001}. This result indicated a very dynamic and
disordered physical picture for the air/water interface.

Femtosecond pump-probe experimental studies have shown that
orientational relaxation dynamics of water molecules in bulk water
is as fast as 1ps, and also the motion of non-hydrogen bonded
water molecules is faster than hydrogen bonded water molecules
\cite{BakkerScience}. Therefore, it is reasonable to predict very
fast orientational motion of interfacial water molecules, which is
less hydrogen bonded as bulk water molecules. However, theoretical
simulation by Chandra \textit{et al.} argued for slower dynamics
for interfacial water molecules than that of bulk ones (10.5ps vs.
7.1ps) \cite{ChandraCPL20032004}. Simulation by Benjamin
\textit{et al.} also suggested that the lifetime of hydrogen bond
at interface between water and an organic liquid is significantly
longer than that of bulk water molecules \cite{BenjaminJPCBASAP}.
No matter how, these all suggested ultrafast orientational motion
of interfacial water molecules. Most importantly, ultrafast
orientational motion may have effects on vibrational spectral
width, as well as the strength of vibrational resonance in
different input/output polarization combinations, provided that
the motion is faster than the time scale of $1/\Gamma_{q}$ (the
rapid motion limit) and the motion covers a broad range centered
around the surface normal, as reported by Wei \textit{et al.}
\cite{WeiXingPRL2001}.

Recent quantitative analysis of data in SFG vibrational
spectroscopy have suggested that vapor/liquid interface are
generally well ordered, and sometimes even with anti-parallel
double-layered structures
\cite{ShenMirandaJPCBReview,ShenLinJCPAcetone2001,TyrodeJPCBpaper,ChenhuaPapers}.
It has been generally accepted that liquid interface with strong
hydrogen bonding between molecules should be well ordered
\cite{ShenMirandaJPCBReview}. Therefore, it is surprising to
comprehend the picture dictated by ultrafast orientational motion
in such a broad angular range.

Here we show from SFG-VS measurements in four sets of experimental
configurations that Wei \textit{et al.}'s analysis
\cite{WeiXingPRL2001} can only be considered valid for the
particular SFG experimental configuration. We conclude that the
orientational motion of water molecules at the air/water interface
can only be libratory within a limited angular range around a
tilting angle around $33^{\circ}$, instead of being around the
interface normal. Therefore, the air/water interface is much
better ordered than previously suggested.

\begin{figure}[h!]
\begin{center}
\includegraphics[height=8cm,width=8cm]{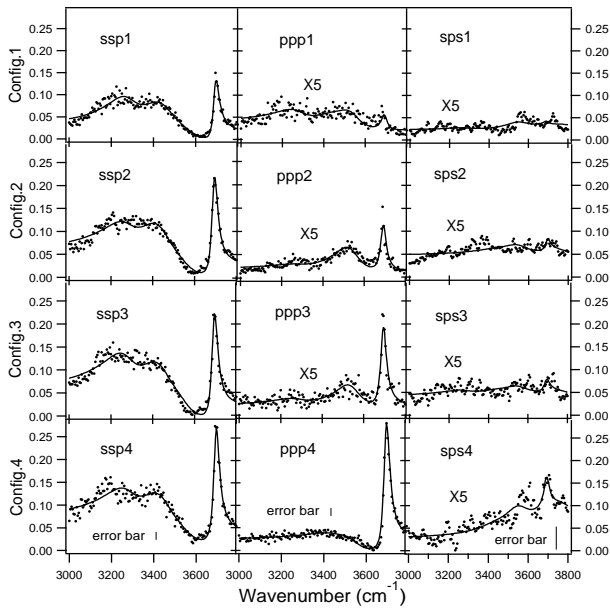}
\caption{SFG spectra of air/water interface in different
polarization combination and experimental configurations. All
spectra are normalized to the same scale. The solid lines are
globally fitted curves with Lorentzian line shape function. Note
the different error bars for graphs in different scales.
}\label{allSpectra}
\end{center}
\end{figure}

The detail of the SFG-VS experiment was described elsewhere
\cite{WaterLongDynamics}. Fig. \ref{allSpectra} shows SFG spectra
of the air/water interface in reflective geometry with
\textit{ssp}, \textit{ppp} and \textit{sps} polarization
combinations at four sets of incident angles for the visible(Vis)
and IR laser beams, namely, Config.1: Vis$=39^{\circ}$,
IR$=55^{\circ}$; Config.2: Vis$=45^{\circ}$, IR=$55^{\circ}$;
Config.3: Vis$=48^{\circ}$, IR$=57^{\circ}$; Config.4:
Vis$=63^{\circ}$, IR$=55^{\circ}$. It is clear that there is
strong dependence of \textit{ppp} and \textit{sps} spectra on the
incident angles, especially on the visible beam. The \textit{sps}
spectra in Config.4 shows clearly much larger intensity at $\sim
3700cm^{-1}$ than those of the other three configurations. Even
though the \textit{sps} intensity is generally much smaller than
that of \textit{ssp} and \textit{ppp} polarizations, the
\textit{sps} intensity is well above the experimental error, and
have been repeated more than a dozen times. It is important to see
that the spectra in Config.2 resembles closely the data obtained
by Wei \textit{et al.} with Vis$=45^{\circ}$, IR=$57^{\circ}$
\cite{WeiXingPRL2001}.

The \textit{ssp} spectra of all four experimental configurations
overlap with each other well within experimental error when
normalized to the $\sim 3700cm^{-1}$ peak intensity. The
difference of their absolute intensities can be quantitatively
accounted from the Fresnel coefficients with different incident
angles. These facts indicate that SFG data with different incident
angles are from interfacial layers without detectable bulk
contributions \cite{ShenApplyPhys}.

Quantitative polarization and orientational analysis in different
experimental configurations can provide detailed information on
structure, conformation and dynamics of molecules at liquid
interfaces
\cite{Shen-5CT,WeiXingPRL2001,WHFRaoJCP2003,Lurong23HongfeiIRPC,HongfeiCJCPPaper}.
The polarization dependent SFG intensity is \cite{Shen-5CT}:

\begin{eqnarray}
I(\omega)&=&\frac{{8\pi ^3 \omega ^2sec^2\beta
}}{{c^{3}n_{1}(\omega)n_{1}(\omega_{1}})n_{1}(\omega_{2})}\left|\chi
_{eff}^{(2)}\right|^2 I(\omega_{1})I(\omega_{2})\label{all}
\end{eqnarray}
\noindent

with

\begin{eqnarray}
\chi^{(2)}_{eff}&=&[\hat{e}(\omega) \cdot \textbf{L}(\omega)]
\cdot \chi^{(2)}_{ijk} : [\textbf{L}(\omega_1)\cdot
\hat{e}(\omega_1)][\textbf{L}(\omega_2) \cdot \hat{e}(\omega_2)]\nonumber\\
\label{vectors}
\end{eqnarray}

\noindent in which $n_{i}(\omega_{i})$ is the refractive index of
phase $i$ at frequency $\omega_{i}$; $\beta$ is the angle of the
laser beam; $I$ is the intensity of the laser beam or the SFG
signal. $\chi_{eff}^{(2)}$ is the effective second order
susceptibility; $\chi^{(2)}_{ijk}$ is the macroscopic
susceptibility tensor elements, determined by the macroscopic and
molecular symmetry \cite{Lurong23HongfeiIRPC};
$\textbf{L}(\omega_{i})$ the local field factor tensor, and
$\hat{e}(\omega_{i})$ the unit vector of the corresponding optical
field, which is responsible for the experimental configuration
dependence.

As Wei \textit{et al.} had demonstrated \cite{WeiXingPRL2001},
when orientational motion average is considered, if the
orientational motion is faster than the vibrational relaxation
time scale $1/\Gamma_{q}$ of the $q$th mode, slow motion average
of the orientation motion is no longer valid. The slow motion
limit gives,

\begin{eqnarray}
\chi^{(2)}_{ijk}&=&N_{s}\sum_{q}\sum_{\lambda\mu\nu}
\frac{a_{q,\lambda\mu\nu}}{\omega_{2}-\omega_{q}+i\Gamma_{q}}
\langle D_{i\lambda}D_{j\mu}D_{k\nu}\rangle \label{SlowAverage}
\end{eqnarray}

\noindent and the fast motion gives,

\begin{eqnarray}
\chi^{(2)}_{ijk}&=&N_{s}\sum_{q}\sum_{\lambda\mu\nu}
\frac{a_{q,\lambda\mu\nu}}{\omega_{2}-\omega_{q}+i\Gamma_{q}}
\langle D_{i\lambda}D_{j\mu}\rangle \langle D_{k\nu}
\rangle\label{FastAverage}
\end{eqnarray}

\noindent in which $N_{s}$ is the surface density of molecules;
$a_{q,\lambda\mu\nu}$ and $\omega_{q}$ and $\Gamma_{q}$ are the
amplitude, resonant frequency and damping constant of the $q$th
molecular vibrational mode, respectively; and
$D_{l\xi}(t)=\hat{l}\cdot\hat{\xi}(t)$ is the time-dependent
direction cosine matrix with $l=i,j,k$ for laboratory coordinates
and $\xi=\lambda,\mu,\nu$ for molecular coordinates.

Wei \textit{et al.} used the step orientational distribution
function in Eq. \ref{StepAverage}, as well as other distribution
functions, such as Gaussian, centered at the surface normal, and
concluded for a fast motion of the interfacial free OH bond in a
range of $\theta_{M}=51^{\circ}$ from the data obtained with
Vis$=45^{\circ}$, IR=$57^{\circ}$ \cite{WeiXingPRL2001}.
Otherwise, the \textit{ppp} and \textit{sps} intensities should be
comparable, given the \textit{ssp} intensity is about 10 times of
that for \textit{ppp} in their SFG data.

\begin{eqnarray}
f(\theta)&=&cost\ \ \ \ for \ \ \ 0\leq \theta
\leq\theta_{M}\nonumber\\
f(\theta)&=&0\ \ \ \ \ \ \ \ for \ \ \ \theta \geq
\theta_{M}\label{StepAverage}
\end{eqnarray}

\begin{figure}[h!]
\begin{center}
\includegraphics[width=6cm]{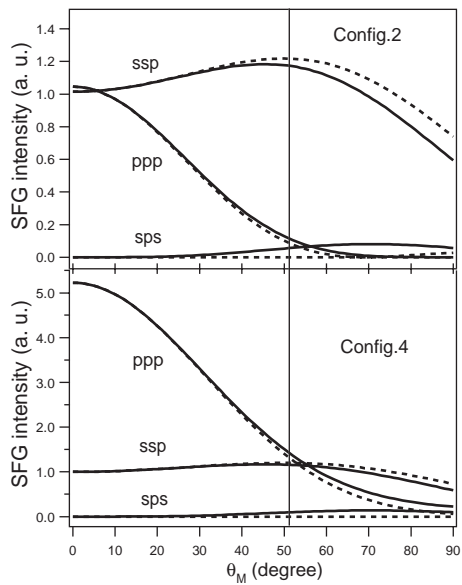}
\caption{SFG intensity of the free OH bond simulated with both
slow motion limit (solid curves) and fast motion limit (dotted
curves) following the procedure and parameters as Wei \textit{et
al.} \cite{WeiXingPRL2001}. In order to compare values in
different configuration, the factor of $\sec^{2}\beta$ is included
in the calculation, and all intensities are normalized to the
\textit{ssp} intensity in Config.4 when $\theta_{M}=0^{\circ}$.
The vertical line indicates the distribution width suggested by
Wei \textit{et al.}}\label{MotionSimulation}
\end{center}
\end{figure}

It is clear that such treatment can not explain the much stronger
\textit{sps} intensity around $\sim 3700cm^{-1}$ in Config.4, even
though it can seemingly explain data in Config.1,2, and 3. Using
the same parameters and calculation procedures as Wei \textit{et
al}, we simulated the \textit{ssp}, \textit{ppp}, and \textit{sps}
intensities with both slow motion and fast motion limit. Fig.
\ref{MotionSimulation} shows the results for Config.2 and
Config.4. Our calculation of Config.2 gives the same results as
that by Wei \textit{et al.} \cite{WeiXingPRL2001}. We noticed in
the calculation that the fast orientational motion centered at
interfacial normal shall make all SFG intensity in \textit{sps}
spectra vanish for all experimental configurations, independent of
the range of the motion. This is clearly in contradiction to the
data from Config.4. As Wei \textit{et al.} concluded, the slow
motion limit certainly can not explain the relative intensity of
the 3700$cm^{-1}$ peak for the \textit{ssp}, \textit{ppp}
polarizations. Therefore, alternative description of the motion
and orientation has to be invoked.

Now we consider the case when $f(\theta)$ is a Gaussian function
around $\theta_{0}$ instead of the interfacial normal.

\begin{eqnarray}
f(\theta)&=&\frac{1}{\sqrt{2\pi\sigma^{2}}}e^{-(\theta-\theta_{0})^{2}/2\sigma^{2}}
\label{Gaussian}
\end{eqnarray}

\begin{figure}[h!]
\begin{center}
\includegraphics[width=6cm]{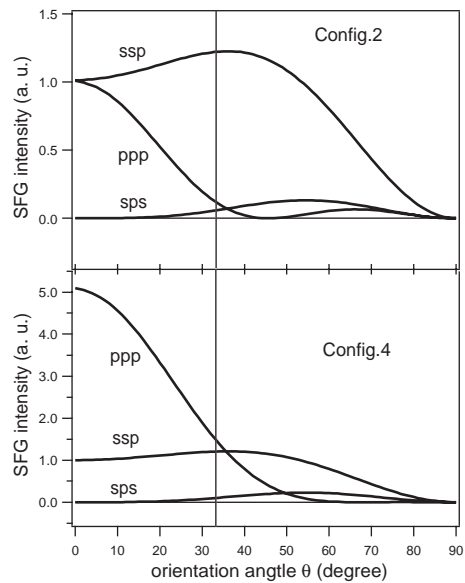}
\caption{Simulated SFG intensity of the free OH bonds at different
tilt angle $\theta_{0}$ with $\sigma=0^{\circ}$. All curves are
normalized to the \textit{ssp} intensity in Config.4 with
$\theta=0^{\circ}$. The vertical line indicates the orientation
which quantitatively explains the observed SFG data. Please note
the difference of the scales for the two configurations.
}\label{C3vSimulation}
\end{center}
\end{figure}

Fig. \ref{C3vSimulation} shows simulation of Config.2 and Config.4
assuming a rotationally isotropic interface with $\sigma=0$, using
the same set of parameters as Wei \textit{et al.}
\cite{WeiXingPRL2001}, and the average procedures
\cite{WeiXingPRL2001,GanweiCPLNull,ChenhuaPapers,Lurong23HongfeiIRPC}.
For data in all four configurations in Fig. \ref{allSpectra}, a
consistent $\theta_{0}=33^{\circ}$ can be reached from
polarization and orientation analysis \cite{WaterLongDynamics}.
This result is in good agreement with the $<38^{\circ}$ tilt angle
reported by Q. Du \textit{et al.} \cite{DuQuanPRL1993}.

It is easy to show that when $\sigma=0$, fast motion and slow
motion treatment would converge. Further simulations show that in
order to satisfy the intensities of around $3700cm^{-1}$ in all
four sets of data, $\sigma$ can not exceed $15^{\circ}$ and
$\theta_{0}$ changes from $33^{\circ}$ to $30^{\circ}$ as $\sigma$
increases. Within such a relatively small range of $\sigma$,
simulation results for fast and slow motions become
indistinguishable. Using step function of $f(\theta)$ centered at
$\theta_{0}$ does not change our general conclusion. This clearly
indicates that the assumption of $f(\theta)$ centered at
interfacial normal is out of the question.

Libration dynamics of the hydrogen bond in liquid water can be as
fast as 0.1ps \cite{FeckoScience2003,Chandler1996PRL}. Therefore,
from the observed SFG data our general conclusion is that such
extremely fast libration dynamics of the tilted free OH bond at
the air/water interface is only possible for orientational motion
within a relatively narrow range.

In the SFG literatures on air/water interface, data were usually
presented for the \textit{ssp} polarization, except for very few
cases \cite{WeiXingPRL2001,TyrodeJPCBpaper}. Polarization analysis
were only performed by Wei \textit{et al.} Our simulation shows
strong dependence for the \textit{ppp} intensity of the free OH
bond spectra on the visible incident angle, especially when around
Vis$=45^{\circ}$. In this range, one or two degree change of the
visible incident angle can cause significant percentage change of
the \textit{ppp} intensity, as can be indicated from data in
Config 2 and 3. Wei \textit{et al.}'s SFG experiment was performed
with Vis$=45^{\circ}$. It happened to be that their analysis with
ultrafast motion in a broad range of orientation centered around
interface normal could just well explain this set of data.
Therefore, they did not consider other possible physical pictures
\cite{ShenPrivate}. Besides, Richmond \textit{et al.} suggested
two possible qualitative explanations for the lack of \textit{sps}
intensity of the free OH stretching mode in the SFG measurement
\cite{richmondJPCB2003p546paper}.

Polarization analysis can further determine the symmetry property
of SFG spectral features \cite{Lurong23HongfeiIRPC}. Analysis
indicates that both the sharp peak at $\sim3700cm^{-1}$ and the
broad peak at $\sim 3550cm^{-1}$ belong to $C_{\infty v}$
symmetry; while both the hydrogen bonded broad peak $\sim
3250cm^{-1}$ and $\sim 3450cm^{-1}$ belong to symmetric type of
the $C_{2v}$ symmetry. Analysis also indicates that the existence
of interfacial water molecules with two free OH bonds is not
supported in SFG data \cite{WaterLongDynamics}. Such interfacial
water molecule shall have $C_{2v}$ symmetry, which should have
shown clear \textit{ssp} spectral intensity around $3650cm^{-1}$
or much stronger \textit{ppp} spectral intensity at the asymmetric
stretching mode position of $3750cm^{-1}$ in at least one of the
four sets of SFG data \cite{Richmond-jpca2000,Mundy-science}.

Thus, $\sim 3550cm^{-1}$ peak with $C_{\infty v}$ symmetry, which
is apparent in the \textit{ppp} polarization of Config.2 and
Config.3, can be assigned to the hydrogen bonded OH of the
interfacial water molecule with a free OH bond. This is also
supported with the fact that $3550cm^{-1}$ is also the frequency
of the hydrogen bonded OH in water dimer \cite{ShenLeePaper}.
Orientational analysis of its peak intensities in different
experimental configurations indicates that this OH bond oriented
with an angle around $\sim140^{\circ}$ from the interface normal.
Indeed, with the free OH pointing to the vapor phase around
$35^{\circ}$, the hydrogen bonded OH of the same molecule pointing
into the liquid phase has to assume an orientation around
$\sim140^{\circ}$. This is consistent with the conclusion that the
dipole vector of the interfacial water molecule is close to
parallel to the interface
\cite{Mundy-science,HynesCP2000JPCB,RiceJCP1991,JPCB2005Tobias}.
The detail of this analysis shall be presented elsewhere.

In summary, quantitative analysis of the SFG spectra in different
polarization and experimental configurations for the air/water
interface shows that orientational motion of the interfacial water
molecule is libratory, as fast as 0.1ps it might be
\cite{FeckoScience2003,Chandler1996PRL}, only within a limited
angular range of less than $15^{\circ}$ with the tilt angle around
$30^{\circ}$. Therefore, the air/water interface is quite
well-ordered. This picture is significantly different from the
previous conclusion that the interfacial water molecule
orientation varies over a broad range within the vibrational
relaxation time, by Wei \textit{et al.} using single SFG
experimental configuration \cite{WeiXingPRL2001}. This progress
provided a direct and detailed physical picture of the orientation
and motion for the air/water interface.

This work was supported by Chines Academy of Sciences (No.
CMS-CX200305), National Natural Science Foundation of China (NSFC
No.20425309) and Chinese Ministry of Science and Technology (MOST
No. G1999075305). H.F.W. acknowledges Y. R. Shen for helpful
discussions.

\end{document}